\documentclass[reprint,twocolumn,floats,english,jcp,aip,preprintnumbers,amsmath,amssymb,array]{revtex4-1}
\usepackage[T1]{fontenc}
\usepackage[latin9]{inputenc}
\usepackage{verbatim}
\usepackage{float}
\usepackage{graphicx}

\newcommand\dd[2][]{\ensuremath{\mathrm{d}^{{#1}}{{#2}}
                                 \thinspace}} 
\newcommand\diffop[2][]{\ensuremath{\frac{\dd[#1]{}}
                                         {\dd{#2^{#1}}}}}  
\def\mat#1{\mathbf{#1}} 
\def\op#1{\hat{#1}} 
\def\bra#1{\left\langle #1 \right|} 
\def\ket#1{\left| #1 \right\rangle} 
\def\BraKet#1#2#3{\left\langle #1 \middle| #2 \middle| #3 \right\rangle}
\def\adaga#1#2{\ensuremath{\op{a}^\dag_{#1}\op{a}_{#2}}}

\makeatletter
\@ifundefined{textcolor}{}
{%
 \definecolor{BLACK}{gray}{0}
 \definecolor{WHITE}{gray}{1}
 \definecolor{RED}{rgb}{1,0,0}
 \definecolor{GREEN}{rgb}{0,1,0}
 \definecolor{BLUE}{rgb}{0,0,1}
 \definecolor{CYAN}{cmyk}{1,0,0,0}
 \definecolor{MAGENTA}{cmyk}{0,1,0,0}
 \definecolor{YELLOW}{cmyk}{0,0,1,0}
 }

\usepackage{dcolumn}
\usepackage{bm}
\usepackage{enumerate}

\makeatother

\usepackage{babel}

\begin{document}

\title{Application of a semiclassical model for the second-quantized
  many-electron Hamiltonian to nonequilibrium quantum transport: The
  resonant level model}

\author{David W.H. Swenson$^{1}$, Tal Levy$^{2}$, Guy Cohen$^{2}$, Eran
Rabani${^2}$ and William H. Miller
} 

\address{${^1}$ Department of Chemistry and Kenneth S. Pitzer Center for
  Theoretical Chemistry, University of California, Berkeley,
  California 94720-1460, and Chemical Sciences Division, Lawrence
  Berkeley National Laboratory, Berkeley, California 94720-1460 {{~\\}}
  ${^2}$ School of Chemistry, The Sackler Faculty of Exact
  Sciences, Tel Aviv University, Tel Aviv 69978, Israel}

\date{\today}

\begin{abstract}
A semiclassical (SC) approach is developed for nonequilibrium quantum
transport in molecular junctions. Following the early work of Miller
and White [{\it J. Chem. Phys.} {\bf 84}, 5059 (1986)], the
many-electron Hamiltonian in second quantization is mapped onto a
classical model that preserves the fermionic character of electrons. The
resulting classical electronic Hamiltonian allows for real-time molecular
dynamics simulations of the many-body problem from an uncorrelated initial 
state to the steady state.  Comparisons with exact results generated for the
resonant level model reveal that a semiclassical treatment of transport
provides a quantitative description of the dynamics at all relevant
timescales for a wide range of bias and gate potentials, and for
different temperatures.  The approach opens a door to treating
nontrivial quantum transport problems that remain far from the reach
of fully quantum methodologies.
\end{abstract}
\maketitle

\section{Introduction}
\label{sec:introduction}
Molecular electronics~\cite{Nitzan2003} has provided means to study
the dynamics of open quantum systems, in which one considers a small,
strongly interacting and highly correlated region (the molecule and
its closest vicinity) coupled to several large, noninteracting baths
(representing the fermionic leads and environment).  While the
equilibrium nature of quantum dynamics in condensed phases has been
mostly resolved,\cite{Leggett87} the intrinsic nonequilibrium nature
of transport through molecular junctions, along with the necessity to
treat fermionic degrees of freedom, poses a much greater theoretical
challenge, and thus remains poorly understood.

Several different paths have been taken to improve the standard
Landauer-B{\"u}ttiker approach~\cite{Landauer70,Buttiker1986} and its
generalization to the multichannel case,\cite{Langreth81} in order to
account for electron-electron and electron-phonon correlations in
molecular junctions. The different approaches can be classified as
perturbative treatments, among which the most notable examples use the
nonequilibrium Green's function (NEGF)
formalism,\cite{Jauho_book,Datta_book} and numerically exact
techniques, perhaps the most prominent of which have been
time-dependent numerical renormalization group
techniques~\cite{White92,Schmitteckert04,Anders05} and the promising
diagrammatic approaches based on path integral
formulations.\cite{Rabani2008,Weiss08,Werner09,Schiro09,Segal10,Cohen11}

These approaches have been applied to a variety of physically
interesting problems including the description of Coulomb and
Franck-Condon blockade,\cite{Averin86,Beenakker91,Koch05} the
nonequilibrium Kondo problem,\cite{Meir92,Werner10} and inelastic
electron tunneling.\cite{Nitzan07,Rabani2008} While successful to a
large extent, these approaches suffer from several limitations, 
including difficulties that arise in the treatment of more complex
environments or when inelastic scattering is governed by interactions
of electrons with soft modes that are dominated by large
anharmonicities.

Parallel to these developments, a completely different paradigm has
been devised based on semiclassical approaches.  These have provided a
useful tool to simulate the dynamics of molecular subsystems coupled
to a fluctuating environment with significant anharmonicities. The most
appealing semiclassical treatments have relied on the mixed
quantum-classical approach~\cite{Tully90,Webster91a,Coker95,Kapral06}
and on the semiclassical initial value
representation.\cite{Miller2001,Miller06,Makri04} The latter have been
applied to a variety of physically interesting condensed phase
problems with remarkable
success.\cite{Miller98a,Makri98,Miller99,Makri99,Rabani99c,Miller00,Miller01b,Makri03,Makri05,Pollak07,Miller08a,Pollak08}
It is interesting to note that such approaches have not received any
attention in the context of nonequilibrium quantum transport, despite
being exact in the harmonic boson
case.\cite{Miller2001,Rabani99r,Thoss2004}

A major goal of the present work is to show how a semiclassical (SC)
approach can treat the dynamics of a many-body quantum system driven
away from equilibrium by the application of a bias voltage. We
describe the transport problem in second quantization, partitioning
the space into an interacting region describing the molecule and its
closest vicinity and a noninteracting region representing the leads
and the environment.  The approach is based on an SC model for the
general second-quantized many-electron Hamiltonian by Miller and White
(MW),\cite{Miller86a} which followed earlier work of McCurdy, Meyer,
and Miller (MMM)~\cite{MillerMcCurdyClassicalElec,Miller79a,SMM,SMMvsCPP} on
constructing classical models for electronic degrees of freedom.  The
essence of MW's model is that each fermionic degree of freedom (i.e., each
pair of one-particle annihilation/creation operators) is described by a
classical degree of freedom (pair of action-angle variables), while in MMM's
earlier work each electronic \emph{state} is described by a classical degree
of freedom.  MW's model is thus a more ``microscopic'' description of the
electronic degrees of freedom, and more importantly much more ``efficient''
(i.e., involving many fewer classical degrees of freedom) for many electron
systems (where the number of electronic states can be much larger than the
number of one-particle annihilation/creation operators).

To assess the accuracy of the proposed approach, we focus on the
resonant level model and derive working expressions to simulate the
left, right and total current using classical trajectories with
quasi-classical initial conditions. Remarkably excellent agreement
in comparison to exact results is achieved for a wide range of bias
and gate voltages and for different temperatures. Our approach
provides a natural framework to study more complex molecular transport
problems with promising performance.

The paper is organized as follows. In Section~\ref{sec:mapping} we summarize
the semiclassical procedure for constructing a classical model of a second
quantized Hamiltonian. Section~\ref{sec:model} describes the resonant
level model, which is used as a test case to assess the accuracy of the
semiclassical approach. Exact quantum mechanical results for the resonant
level model are provided in Section~\ref{sec:exact}.
Section~\ref{sec:results} summarizes the main results and provides a
detailed comparison between the semiclassical approach and the exact quantum
mechanical treatment. The comparisons cover a wide range of gate and bias
potentials from high to low temperatures.  Section~\ref{sec:conclusions}
summarizes and concludes.

\section{Semiclassical Approach}
\label{sec:mapping}
\subsection{Mapping}
The approach that has been used~\cite{Miller86a,Miller79a,SMM,SMMvsCPP} to
construct semiclassical models for electronic degrees of freedom is to
invert (as meaningfully as possible) the Heisenberg correspondence relation,
\begin{equation}
  \BraKet{n^\prime}{\op{A}}{n} = (2\pi)^{-F} \int_0^{2\pi}
  e^{-i(n^\prime-n)q} A_\textrm{cl}(\bar{n},q) \dd{q}
  \label{fermion:eq:HeisenbergCorrespondence}
\end{equation}
where $\bar{n} = \frac{1}{2}(n^\prime+n)$, $F$ is the number of degrees
of freedom, and $A_\textrm{cl}(\bar{n}, q)$ is a function of the classical
action-angle variables. The Heisenberg correpondence relation was originally
used to obtain approximate matrix elements for $\op{A}$ from the
corresponding classical function of action-angle variables; the
semiclassical goal here is to obtain a classical function of action-angle
variables that corresponds (as best as possible) to the given quantum
mechanical matrix elements of operator $\op{A}$. The formal inverse of the
Fourier transform in Eq.~\eqref{fermion:eq:HeisenbergCorrespondence}
gives the angle dependence of the classical function as a Fourier series,
\begin{equation}
  A_\textrm{cl}(\bar{n},q) = \sum_{k\in \mathbb{Z}^F} e^{i k q}
  \BraKet{n + \frac{k}{2}}{\op{A}}{n-\frac{k}{2}}
  \label{fermion:eq:invHeisenbergCorr}.
\end{equation}
Here, however, the procedure becomes nonunique (and nonexact) because the
classical action variable $\bar{n}$ has to interpolate for all real numbers,
while the matrix elements are only defined for integer values of
$n\pm\frac{k}{2}$.  

One way to proceed is to use the spin-matrix mapping (SMM) method of Meyer
and Miller (MM),\cite{SMM,SMMvsCPP} which utilizes the fact that a general
two-state system is equivalent to a spin $\frac{1}{2}$ system. Thus any
$2\times 2$ matrix $A_{n,n^\prime}$, $n,n^\prime\in\{0,1\}$, can be written
as a linear combination of the three spin matrices $\mat{S}_x$, $\mat{S}_y$,
and $\mat{S}_z$ (and the $2\times 2$ identity matrix). One then uses the
classical expressions for the spin angular momentum,
\begin{subequations}
\begin{align}
  S_x &= \sqrt{\sigma^2 - m^2} \cos(q) 
  \label{fermion:eq:Sx_cl} \\
  S_y &= \sqrt{\sigma^2 - m^2} \sin(q) 
  \label{fermion:eq:Sy_cl} \\
  S_z &= m .
\end{align}
\end{subequations}
in terms of the action-angle variables $(m,q)$. The quantum values of the
action variable $m$ (the projection quantum number) are $\pm\frac{1}{2}$,
but it is convenient to have the quantum values of the action variables be
$0$ (unoccupied) and $1$ (occupied). The elementary canonical transformation
$n=m+\frac{1}{2}$ makes this change, giving the classical function of
action-angles $(n,q)$ for a general $2\times 2$ matrix $A$ as
\begin{align}
  A(n,q) &= (1-n)\,A_{00} + n\,A_{11} \notag \\
  &\qquad 
  +\sqrt{n-n^2+\lambda} \left( A_{10}\, e^{iq} + A_{01}\,e^{-iq} \right)
\label{fermion:eq:general_SMM}
\end{align}
where $\lambda = \sigma^2 - \frac{1}{4}$. The value of $\sigma^2$ will be
discussed below.

To employ this approach on a given second-quantized Hamiltonian, MW used
this SMM model for each term in a Hamiltonian. The Hamiltonian in the
present work has terms of the form $\op{a}^\dag_i\op{a}_i$ and
$\op{a}^\dag_i\op{a}_j$.  The matrix elements of the operator
$\op{a}^\dag_i\op{a}_i$ are diagonal,
\begin{subequations}
\begin{equation}
  \BraKet{n^\prime}{\op{a}^\dag_i\op{a}_i}{n} = n_i
  \prod_j\delta_{n^\prime_j,n_j}
  ,
\end{equation}
which corresponds to Eq.~\eqref{fermion:eq:general_SMM} with $A_{00}=0$ and
$A_{11}=1$ for the $i$th degree of freedom, and to the identity matrix for
all other degrees of freedom. The SMM model thus gives
\begin{equation}
  \op{a}^\dag_i \op{a}_i \mapsto = n_i
  .
\end{equation}
\end{subequations}

The matrix elements for $\op{a}^\dag_i\op{a}_j$ are
\begin{subequations}
\begin{align}
  \BraKet{n^\prime}{\adaga{i}{j}}{n} &= 
  \delta_{n_i^\prime, n_i+1} \delta_{n_j^\prime, n_j-1} \notag \\
  &\quad \times
  \prod_{k\ne i,j} \delta_{n_k^\prime, n_k}  
  \prod_{p=i+1}^{j-1} \left( -1\right)^{n_p} 
  \label{fermion:eq:adagiaj_matrixelem}
\end{align}
where we have assumed that $i<j$. In the case $i>j$, the product over $p$
goes from $j+1$ to $i-1$: it always includes the states between (but not
including) $i$ and $j$. That product is the result of the anticommutation
relation of fermionic creation/annihilation operators.

These matrix elements are separable products of $2\times 2$ matrices: for
the $i$th degree of freedom (corresponding to the factor
$\delta_{n^\prime_i, n^\prime_i+1}$) the $2\times 2$ matrix has $A_{10} = 1$
as the only nonzero matrix element, so that
Eq.~\eqref{fermion:eq:general_SMM} gives the corresponding classical
function as
\begin{equation}
  \sqrt{n_i-n_i^2+\lambda}\,e^{iq_i}.
\end{equation}
For the $j$th degree of freedom (i.e., the factor
$\delta_{n^\prime_j,n_j-1}$), the only nonzero matrix element is $A_{01}=1$,
resulting in the classical function
\begin{equation}
  \sqrt{n_j-n_j^2+\lambda}\,e^{-iq_j}.
\end{equation}
For each degree of freedom $p$, with $p$ between $i$ and $j$ (in normal
ordered form), the factor $\delta_{n^\prime_p,n_p}(-1)^{n_p}$ from
Eq.~\eqref{fermion:eq:adagiaj_matrixelem} corresponds to a diagonal $2\times
2$ matrix in Eq.~\eqref{fermion:eq:general_SMM} with $A_{00} = 1$ and
$A_{11} = -1$. This SMM method thus gives the classical function
\begin{equation}
  1-2n_p.
\end{equation}
\end{subequations}
The net result is
\begin{align}
  \op{a}^\dag_i\op{a}_j &\mapsto \sqrt{\left( n_i-n_i^2+\lambda \right)
  \left( n_j-n_j^2+\lambda \right)}\, e^{i(q_i-q_j)}\notag \\
  &\qquad \times\prod_{p=i}^j f_b(n_p)
  \label{fermion:eq:adagiaj}
\end{align}
where the SMM method gives $f_b(n_p) = 1-2n_p$. For the present application,
however, we found that using these factors $f_b(n_p)$ significantly
underestimates the current (see section~\ref{sec:results}), while excellent
results were obtained by omitting these factors, i.e., by setting
$f_b(n_p)=1$. We also tried an alternative, $f_b(n_p) = \exp(i\pi n_p)$, but
it had the same defect. The issue seems to be that with the quantum values
of $n_p\in\{0,1\}$, the product of these factors is $\pm1$, while the
classical value of $n_p$ frequently leads to $f_b(n_p) < 1$. The product of
many such factors leads to factors much less than $1$, and thus off-diagonal
coupling that are much too small.

We now return to the value of $\sigma$ (which determines $\lambda$).
For a classical spin with $s=\frac{1}{2}$, $\sigma^2=s^2=\frac{1}{4}$,
such that $\lambda = 0$. However, MM suggested using the Langer modified
value $\sigma^2 = \left( s+\frac{1}{2} \right)^2 = 1$, giving $\lambda
= \frac{3}{4}$.  Here, we suggest instead using the quantum value of
$\sigma^2 = s(s+1) = \frac{3}{4}$, and thus $\lambda =
\frac{1}{2}$. The results in section~\ref{sec:results} give empirical
reasons for this choice.

The choice $\lambda=\frac{1}{2}$ also can be justified with the following
argument. Because the goal is to generate a mapping that accounts for the
correct dynamics, it is reasonable to try to match the short-time dynamics
of a simple problem. Consider the Hamiltonian $\op{H} = k
\left(\op{a}^\dag_A\op{a}_B + \op{a}^\dag_B\op{a}_A\right)$, with initial
population in state $A$. The first time derivative of the initial ($t=0$)
quantum population of state $A$ is zero, as is that of our semiclassical
model (averaged over angles).  The second derivative of the initial quantum
population of state $A$ is $-2k^2$, while the semiclassical model gives
$-4k^2\lambda$. If these are to be equal, then we must set
$\lambda=\frac{1}{2}$.

To summarize, the modifications we have made to the original MW model are
(1) to use a value of $\lambda=\frac{1}{2}$ rather than the previously
suggested Langer modified value of $\frac{3}{4}$, and (2) to set the factors
$f_b(n_p)$ in equation~\eqref{fermion:eq:adagiaj} to unity. 

\subsection{Intial conditions}
The semiclassical mapping also requires the selection of initial
conditions for the action-angle variables. It is clear that a na\"ive
approach based on the classical mapping described above will not
provide the correct statistical treatment: Even for the one-particle
system $\op{H}=\op{a}^\dag\op{a}\mapsto n$, it does not reproduce the
quantum partition function when the action-angle variables are sampled
from the corresponding classical thermal distribution.

To recover the correct statistical behavior (at least at time $t=0$)
we use a quasi-classical procedure. Since we are interested in a
noncorrelated initial state with thermally populated leads and an
unpopulated quantum dot, we can populate each degree of freedom
independently.  We enforce quantum statistics on the initial conditions for
each degree of freedom $i$ by setting the initial action $n_i$ to
either $0$ or $1$ such that the expectation value of the action
$\left\langle n_i\right\rangle$, averaged over the set of initial
conditions,  satisfies the Fermi distribution, 
\begin{equation}
  f(\epsilon_i-\mu_i) = \left( 1+e^{\beta(\epsilon_i - \mu_i)} \right)^{-1}
  \label{fermion:eq:fermi_dist}
\end{equation}
where $\mu_i$ is the chemical potential of the lead in which mode $i$
is located and $\beta = 1/T$ is the inverse temperature. The angle
variable $q_i$ is selected at random between $0$ and $2\pi$.

\section{Model Hamiltonian}
\label{sec:model}
We use the resonant level (Landauer) model as an example of quantum
transport. It consists of a single quantum dot state coupled to two
electrodes (left and right) according to the Hamiltonian
\begin{equation}
  \op{H} = \epsilon_0 \op{a}^\dag_0\op{a}_0 
  + \sum_{k=1}^N \epsilon_k \op{a}^\dag_k\op{a}_k
  + \sum_{k=1}^{N} t_k \left(\op{a}^\dag_0 \op{a}_k + \op{a}^\dag_k
  \op{a}_0\right),
  \label{fermion:eq:LandauerH}
\end{equation}
where $\epsilon_0$ is the energy of the isolated quantum dot (and will also
be used to model a gate voltage), $\epsilon_k$ is the energy
associated with the electrode mode $k$, and $t_k$ is the coupling between
the quantum dot and the electrode mode $k$.

Using the procedure described in section~\ref{sec:mapping}, we construct the
classical model of this Hamiltonian in action-angle variables:
\begin{align}
  H(n,q) &= \epsilon_0 n_0 + \sum_{k=1}^N \epsilon_k n_k 
  + \sum_{k=1}^N t_k \sqrt{n_0-n_0^2+\frac{1}{2}}
  \notag \\ &\quad \times
  \sqrt{n_k-n_k^2+\frac{1}{2}} \left( e^{i(q_0-q_k)}\prod_{p=1}^N f_b(n_p) +
  \textrm{H.c}\right)
  \label{fermion:eq:LandauerH_SC}
  .
\end{align}

The left current is given by the change in occupancy of the left
electrode, with the right current defined analogously and the total current
given by half the difference of the two. For the Hamiltonian in
Eq.~\eqref{fermion:eq:LandauerH}, the Heisenberg time derivative gives the
left current as:
\begin{align}
  \op{I}_L(t) &= -e \diffop{t}\left\langle\sum_{k\in
  L}\op{a}^\dag_k\op{a}_k\right\rangle \\
  &= -\frac{e\,i}{\hbar} \left\langle\sum_{k\in L}t_k \left(
  \op{a}^\dag_0\op{a}_k - \op{a}^\dag_k\op{a}_0 \right)\right\rangle
  \label{eq:current}
\end{align}
where $L$ is the set of states in the left electrode.

We can choose to take the semiclassical approximation before the time
derivative by mapping the occupation to a classical quantity and taking its
time derivative, or we can take the result of the quantum time derivative
and map that to a classical quantity. For this system, our mapping method
gives formally equivalent results for either procedure.

The electrodes are described within the wide band limit with a sharp
cutoff at high and low energy values:
\begin{equation}
  J_{L/R}(\epsilon) = \frac{\Gamma_{L/R}}{
  \left( 1+e^{A\left(\epsilon-\frac{B}{2}\right)} \right)
  \left( 1+e^{-A\left(\epsilon+\frac{B}{2}\right)} \right)}
  \label{fermion:eq:spectral_density}
\end{equation}
where, in all results reported below we use $\Gamma_L=\Gamma_R =
\frac{1}{2}$, $\Gamma = \Gamma_L+\Gamma_R$, $A=5\Gamma$, and $B=20\Gamma$.
For the semiclassical mapping, we use a uniform discretization to select the
energies of the leads' states, and thus the couplings are given by
\begin{equation}
  t_k(\epsilon_k) = \sqrt{\frac{J(\epsilon_k) \Delta\epsilon}{2\pi}}
  .
\end{equation}

\section{Exact Quantum Mechanics}
\label{sec:exact}
An exact quantum mechanical solution for the transient current for the
Hamiltonian specified above is straightforward to derive, and thus provides
means to assess the accuracy of the semiclassical treatment. Under the
assumptions of no correlation at $t=0$ 
\begin{equation} \left\langle
a_{k}^{\dagger}\left(0\right)a_0\left(0\right)\right\rangle =\left\langle
a_0^{\dagger}\left(0\right)a_{k}\left(0\right)\right\rangle =0,
\end{equation}
and a Boltzmann distribution for the leads' populations
\begin{equation}
\left\langle
a_{k}^{\dagger}\left(0\right)a_{k'}\left(0\right)\right\rangle
=f\left(\epsilon_{k}-\mu_{L,R}\right)\delta_{kk'},
\end{equation}
where $f\left(\epsilon\right)$ is the Fermi function and as before
$\mu_{L,R}$ is the chemical potential for the left (L) or right (R)
lead, one can derive an exact expression for the left current given in
Eq.~(\ref{eq:current}):
\begin{equation}
I_{L}\left(t\right)=\frac{2e}{\hbar}\mbox{Im}\left\{
\frac{1}{2\pi}\int_{-\infty}^{\infty}e^{-i\omega'
  t}J_{L}\left(\omega'\right)\mbox{d}\omega'\right\}
\label{eq:exact-current}
\end{equation}
for any given initial dot population
\begin{equation}
\left\langle a_0^{\dagger}\left(0\right)a_0\left(0\right)\right\rangle
= n_0 \in [0,1].
\end{equation}
In the above, $J_{L}\left(\omega'\right)$ is given by:
\begin{widetext}
\begin{eqnarray}  
  J_{L}\left(\omega'\right) & = &
  \frac{1}{2\pi}\int_{-\infty}^{\infty}\left(\frac{1}{-i\omega'}\frac{\left(-\frac{i}{2}\Sigma_{L}^{<}\left(\omega\right)+\left(-\frac{i}{2}\Sigma_{L}^{<}\left(\omega-\omega'\right)\right)^{\dagger}\right)}{\left(\omega-\frac{\epsilon_0}{\hbar}-\frac{1}{\hbar}\Sigma_{L,R}\left(\omega\right)\right)}\right.\nonumber
  \\ & - &
  \frac{\left(\Sigma_{L}\left(\omega-\omega'\right)\right)^{\dagger}n_{0}\left(0\right)}{\left(\omega-\omega'-\frac{\epsilon_0}{\hbar}-\frac{1}{\hbar}\left(\Sigma_{L,R}\left(\omega-\omega'\right)\right)^{\dagger}\right)\left(\omega-\frac{\epsilon_0}{\hbar}-\frac{1}{\hbar}\Sigma_{L,R}\left(\omega\right)\right)}\nonumber
  \\ & + &
  \left.\frac{1}{-i\omega'\hbar}\frac{\left(\Sigma_{L}\left(\omega-\omega'\right)\right)^{\dagger}\left(-\frac{i}{2}\Sigma_{L,R}^{<}\left(\omega\right)+\left(-\frac{i}{2}\Sigma_{L,R}^{<}\left(\omega-\omega'\right)\right)^{\dagger}\right)}{\left(\omega-\omega'-\frac{\epsilon_0}{\hbar}-\frac{1}{\hbar}\left(\Sigma_{L,R}\left(\omega-\omega'\right)\right)^{\dagger}\right)\left(\omega-\frac{\epsilon_0}{\hbar}-\frac{1}{\hbar}\Sigma_{L,R}\left(\omega\right)\right)}\right)\mbox{d}\omega,
  \label{eq:left}
\end{eqnarray}
\end{widetext}
$\Sigma\left(\omega\right)$ is the self energy, and
$\Sigma^{<}\left(\omega\right)$ is the lesser self energy, both
specified below. A similar expression can be derived for the right
current $\left(I_{R}\left(t\right)\right)$ by the replacement $L
\leftrightarrow R$. The total current is given by the different of the
left and right currents, $I\left(t\right) = \frac{I_{L}\left(t\right)
  - I_{R}\left(t\right)}{2}$.

As a check on the above, we discuss two known limits for the current. In the
limit $t\rightarrow 0$, it is simple to show that the current vanishes. This
is a result of the initial preparation of an uncorrelated state and can be
derived with the help of the initial value theorem
\begin{equation}
\lim_{t\rightarrow0}I_{L}\left(t\right)=\frac{2e}{\hbar}\mbox{Im}\left\{
\lim_{-i\omega'\rightarrow\infty}-i\omega'J\left(\omega'\right)\right\}
=0,
\end{equation}
where we assumed that the self-energies vanish at the boundaries in
the frequency domain:
\begin{equation}
\lim_{-i\omega'\rightarrow\infty}\Sigma\left(\omega\right)=\lim_{-i\omega'\rightarrow\infty}\Sigma^{<}\left(\omega\right)=0.
\end{equation}

To recover the Landauer expression for the current at
steady state~\cite{Landauer70,Buttiker1986} we take the limit
$t\rightarrow\infty$ in Eq.~(\ref{eq:exact-current}), this time with
the aid of the final value theorem
\begin{widetext}
\begin{eqnarray}
\lim_{t\rightarrow\infty}I_{L}\left(t\right) & = &
\frac{2e}{\hbar}\mbox{Im}\left\{
\lim_{-i\omega'\rightarrow0}-i\omega'J\left(\omega'\right)\right\}
\nonumber \\ & = & \frac{2e}{\hbar}\mbox{Im}\left\{
\frac{1}{2\pi}\int_{-\infty}^{\infty}\left(\frac{\left(\frac{i}{2}\Sigma_{L}^{<}\left(\omega\right)+\left(\frac{i}{2}\Sigma_{L}^{<}\left(\omega\right)\right)^{\dagger}\right)}{\left(\omega-\frac{\epsilon_0}{\hbar}-\frac{1}{\hbar}\Sigma_{L,R}\left(\omega\right)\right)}\right.\right.\nonumber
\\ & &
+\left.\left.\frac{1}{\hbar}\frac{\left(\Sigma_{L}\left(\omega\right)\right)^{\dagger}\left(\frac{i}{2}\Sigma_{L,R}^{<}\left(\omega\right)+\left(\frac{i}{2}\Sigma_{L,R}^{<}\left(\omega\right)\right)^{\dagger}\right)}{\left(\omega-\frac{\epsilon_0}{\hbar}-\frac{1}{\hbar}\left(\Sigma_{L,R}\left(\omega\right)\right)^{\dagger}\right)\left(\omega-\frac{\epsilon_0}{\hbar}-\frac{1}{\hbar}\Sigma_{L,R}\left(\omega\right)\right)}\right)\mbox{d}\omega\right\}.
\label{eq:lim-lb}
\end{eqnarray}
\end{widetext}
Rearranging Eq.~(\ref{eq:lim-lb}) and using the well known
representation of the self energies in terms of the real
($\Lambda_{L,R}\left(\omega\right)$) and imaginary
($\Gamma_{L,R}\left(\omega\right)$) portions:
\begin{equation}
  \Sigma_{L,R}\left(\omega\right)=\Lambda_{L,R}\left(\omega\right)-\frac{1}{2}i\Gamma_{L,R}\left(\omega\right)
\end{equation}
\begin{equation}
\Sigma_{L,R}^{<}\left(\omega\right)=if\left(\hbar\omega-\mu_{L,R}\right)
\Gamma_{L,R}\left(\omega\right)
\end{equation}
we finally arrive at the Landauer expression for the
current:
\begin{eqnarray}
\nonumber \lim_{t\rightarrow\infty}I_{L}\left(t\right) & = &
\frac{e}{2\pi\hbar^{2}}\int_{-\infty}^{\infty} 
\left(f(\hbar\omega-\mu_{L})-f(\hbar\omega-\mu_{R})\right) \\ & &
\times \frac{
  \Gamma_{R}\left(\omega\right)\Gamma_{L}\left(\omega\right)}
       {\left(\omega-\frac{\tilde{\epsilon_0}}{\hbar}\right)^{2}+\frac{1}{4\hbar^{2}}\Gamma\left(\omega\right)^{2}} \mbox{d}\omega
\end{eqnarray}
where the rescaled energy $\tilde{\epsilon_0}$ is given by:
\begin{equation}
\tilde{\epsilon_0} = \epsilon_0 + \Lambda_{L}\left(\omega\right) +
\Lambda_{R}\left(\omega\right)
\end{equation}
and $\Gamma\left(\omega\right) = \Gamma_L\left(\omega\right) +
\Gamma_R\left(\omega\right)$.

We choose
\begin{equation}
  \Gamma_{L/R}(\omega) = \frac{\Gamma_{L/R}}{\left(e^{A\left(
  \omega-\frac{B}{2} \right)} + 1\right) \left( e^{-A\left(
  \omega+\frac{B}{2} \right)} + 1 \right)}
\end{equation}
where $\Gamma_L=\Gamma_R=\frac{1}{2}$. $A$ and $B$ are defined in
Eq.~\eqref{fermion:eq:spectral_density}. $\Lambda_{L/R}(\omega)$ is obtained
from the Kramers-Kronig relation
\begin{equation}
  \Lambda_{L/R}(\omega) = \frac{1}{\pi} \text{pp} \int_{-\infty}^\infty
  \dd{\omega^\prime} \frac{\Gamma_{L/R}(\omega^\prime)}{\omega^\prime -
  \omega}
\end{equation}
where $\text{pp}$ denotes the Cauchy principle value.

\section{Results}
\label{sec:results}
Our simulations used the semiclassical Hamiltonian of
Eq.~\eqref{fermion:eq:LandauerH_SC} in action-angle variables, with initial
conditions selected as described in section~\ref{sec:mapping}.  Numerical
integration of the trajectories was performed with the sixth-order Gear
predictor-corrector algorithm, modified to allow adaptive timesteps
(necessary due to the square roots in the Hamiltonian).  The maximum step
size was $\Delta t=0.01\hbar/\Gamma$. We used 400 states per lead and
reported results with $2\times 10^5$ trajectories (see below for further
discussion on these points). A calculation to time $t=8\hbar/\Gamma$ with
400 modes per electrode and $2\times 10^5$ trajectories required less than
one hour walltime with 120 computational cores.

\begin{figure}[t]
\includegraphics[width=8cm]{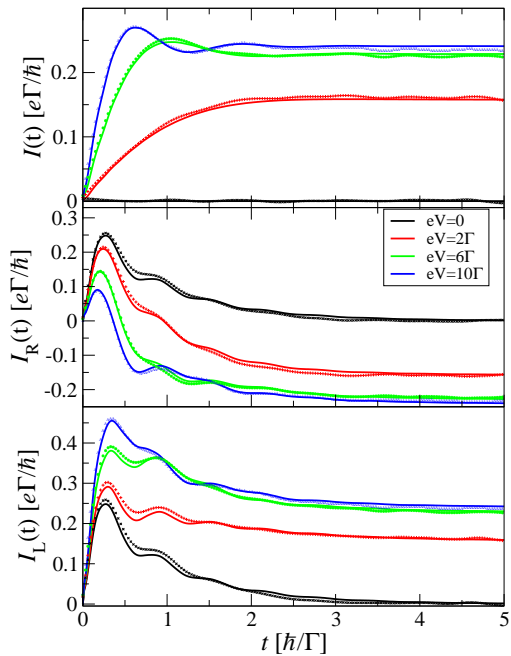}
\caption{Plots of the left (lower panel), right (middle panel) and
  total (upper panel) transient current for different values of the
  source-drain voltage $eV=\mu_L-\mu_R$.  Other model parameters are:
  $T=\Gamma/3$, $\epsilon_0=0$ and $N_L=N_R=400$. Solid line and
  symbols correspond to exact quantum mechanical and semiclassical
  results, respectively.}
\label{fig:mu}
\end{figure}

\begin{figure}[t]
\includegraphics[width=8cm]{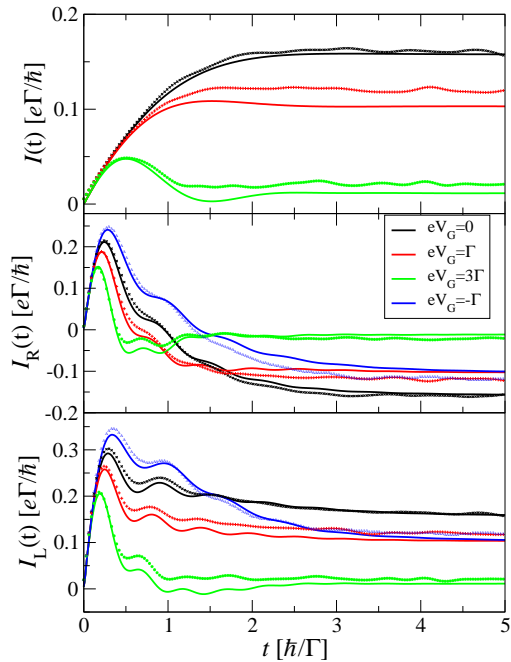}
\caption{Plots of the left (lower panel), right (middle panel) and
  total (upper panel) transient current for different values of the
  gate voltage $eV_g=\epsilon_0$.  Other model parameters are:
  $\mu_L=-\mu_R=\Gamma$, $T=\Gamma/3$ and $N_L=N_R=400$. Solid line
  and symbols correspond to exact quantum mechanical and
  semiclassical results, respectively.}
\label{fig:gate}
\end{figure}

\begin{figure}[t]
\includegraphics[width=8cm]{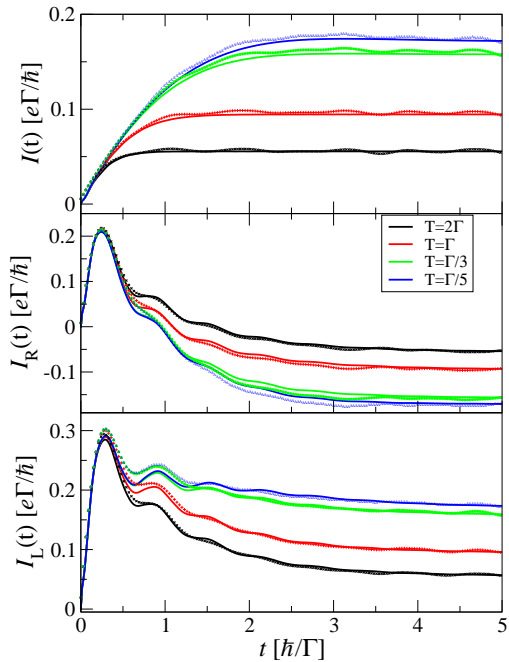}
\caption{Plots of the left (lower panel), right (middle panel) and
  total (upper panel) transient current for different
  temperatures. Other model parameters are: $\mu_L=-\mu_R=\Gamma$,
  $\epsilon_0=0$ and $N_L=N_R=400$.  Solid line and symbols
  correspond to exact quantum mechanical and semiclassical results,
  respectively.}
\label{fig:beta}
\end{figure}

In Figs.~1-3, we plot the left (lower panels), right (middle panels) and
total (upper panels) currents as a function of time for different
source-drain voltages (Fig.~\ref{fig:mu}), gate voltages
(Fig.~\ref{fig:gate}) and temperatures (Fig.~\ref{fig:beta}). The left,
right and total currents show a distinct time-dependence and decay to
the same value at steady state. In these plots, the exact quantum results of
section~\ref{sec:exact} are presented as lines, and the results of the
semiclassical model presented in section~\ref{sec:mapping} are presented as
symbols.

Figure~\ref{fig:mu} shows the time-dependent current at a wide range of
source-drain voltages $eV=\mu_L-\mu_R$, with $\epsilon_0=0$ (zero gate
voltage) and temperature $T=\Gamma/3$. The largest values of this bias are
limited by the width of our band, $B=20\Gamma$ [see
Eq.~\eqref{fermion:eq:spectral_density}]. The most striking result
shown in Fig.~\ref{fig:mu} is the excellent agreement between the
semiclassical method and the exact quantum mechanical result. In some cases
it is difficult to distinguish the two. 
The semiclassical approach captures both the evolution of the current at
early and intermediate times, and its decay to the correct steady-state
value at longer time. The agreement is quantitative for all source-drain
voltages studied. The pronounced oscillations in the left and right current
as they decay to steady state result from the finite band width of the
leads, and are also captured by the semiclassical method.

At zero source-drain bias, the left and right currents show a significant
transient effect until the zero-current steady-state value is reached, while
the total current is identically zero at all times. This suggests that for
other systems the total current would be a better observable since one can
infer the steady-state result from relatively shorter
times.~\cite{Rabani08,Peskin} For the largest bias presented,
$I=0.241\Gamma$, which approaches the infinite bias limit of
$I=0.25\Gamma$.

In Fig.~\ref{fig:gate} we plot the time-dependent current for different
gate voltages, $e V_G$. The gate voltage is determined by the parameter
$e V_G = \epsilon_0$ in the Hamiltonian~\eqref{fermion:eq:LandauerH}. The
line for $e V_G=-\Gamma$ is not plotted for the total current because the
quantum result overlaps the $e V_G=\Gamma$ line (the semiclassical results
for these two series nearly overlap as well). For nonzero gate voltages, we
find that the semiclassical method is not as accurate as for the case where
$eV_G = 0$, although it still captures the correct trends at all times.  In
particular, the even more pronounced oscillations and the decay of the
amplitude of the oscillations are captured by the semiclassical treatment.
Additionally, the steady-state currents for gate voltages of equal magnitude
but opposite signs (e.g., $e V_G = \pm\Gamma$) are equal (within the
numerical noise), just as in the quantum mechanical results for these
parameters.

The effect of temperature is shown in Fig.~\ref{fig:beta}. We
explored a range temperatures from $\Gamma/5$ to $2\Gamma$, which covers
the classical to quantum regimes. We find excellent agreement for all
temperatures, even when considering the left and right currents which
exhibit transient phenomena on longer timescales and are thus more difficult
to describe.~\cite{Rabani08} A closer examination of the results reveals
that the semiclassical treatment is a bit more accurate for higher
temperatures (as might be expected), yet the overall agreement for all
temperatures is quite surprising, in particular in view of the failures of
semiclassical treatments at low temperatures for other
systems.~\cite{Rabani99r} 

\begin{figure}[t]
\includegraphics[width=8cm]{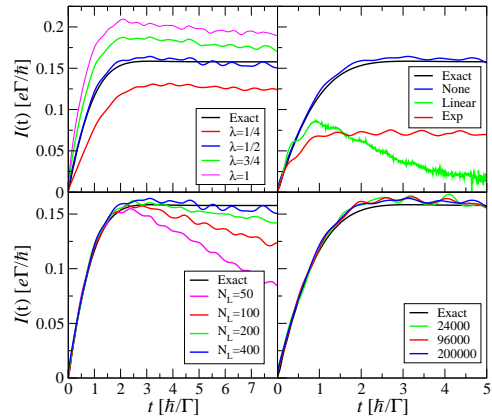}
\caption{Plots of the total transient current for different values of the
Langer modification (upper left panel), for different choices of the factor
$f_b(n)$ (upper right panel), for different numbers of states in the leads
(lower left panel), and with different numbers of trajectories (lower right
panel).  In all panels $\mu_L=-\mu_R=\Gamma$, $\epsilon_0=0$, $T=\Gamma/3$
and $N_L=N_R$.}
\label{fig:modes-lambda}
\end{figure}

Fig.~\ref{fig:modes-lambda} deals with the choices we made in the
semiclassical procedure described in section~\ref{sec:mapping}. The upper
left panel displays the effect of choosing different values of $\lambda$
(or, equivalently, $\sigma^2$). The choice $\lambda=\frac{1}{2}$ corresponds
to the quantum $\sigma^2=s(s+1)$, and is what was used in the results shown
in Figs.~1--3. The choice $\lambda=\frac{3}{4}$ is the the Langer-modified
$\sigma^2=(s+\frac{1}{2})^2$ suggested by MW. Stock has treated the Langer
modification as a free parameter,~\cite{StockLanger} and so in that spirit
we include the choices $\lambda=1$ and $\lambda=\frac{1}{4}$, the latter
being near Stock's result of $0.3125$ (for a different system). In
particular, note the short-time behavior: The slope of the semiclassical
current at time $t=0$ can be shown to be proportional to $\lambda$, and
$\lambda=\frac{1}{2}$ best matches the exact quantum result. This is
consistent with the short-time analysis in section~\ref{sec:mapping}.

The upper right panel of Fig.~\ref{fig:modes-lambda} shows how the function
$f_b(n)$, which relates to the anticommutation relations, affects the
results. The SMM form for these factors, $f_b(n) = 1-2n$, is labeled as
``Linear,'' while the alternate we explored, $f_b(n) = \exp(i\pi n)$, is
labeled ``Exp.'' The method we used in the results presented in Figs.~1--3
was to omit these terms (i.e., $f_b(n)=1$) and is labelled ``None.''
Clearly, the linear and exponential versions underestimate the current (the
linear version tending to zero steady-state current), while omitting these
factors gives extremely good agreement. This can be traced to the fact that
quantum mechanically, the values of these factors are either $-1$ or $1$
which can be absorbed into the coupling constants $t_k$ as a phase factor.
One can prove that the current is independent of the sign of $t_k$ for the
initially uncorrelated case.  However, in the semiclassical mapping the
values of these factors are continuous.  They can therefore significantly
change the effective value of the coupling constants, decreasing the
current. It is also notable that the oscillations in the linear and
exponential versions are more pronounced than when these factors are
omitted.

The lower left panel shows the importance of using a sufficient number of
modes per electrode. Since the semiclassical simulations are performed for a
finite number of states in the electrodes, the ``steady-state'' current we
obtain cannot hold to infinite time. All other results presented involved
400 modes per electrode, which Fig.~\ref{fig:modes-lambda} suggests holds
the steady state until at least time $t\approx5\hbar/\Gamma$.

The lower right panel of Fig.~\ref{fig:modes-lambda} shows how the results
converge with the number of trajectories. Even at $2.4\times 10^{4}$
trajectories, the approximate result is clear. The results presented in
Figs.~1--3 are with $2\times 10^{5}$ trajectories.

\section{Concluding Remarks}
\label{sec:conclusions}
The present classical model for electronic degrees of freedom (dofs) is seen
to provide an excellent description of the dynamics of electron transmission
in a simple model of a molecular transistor driven out of equilibrium. The
near quantitative results, obtained over a wide range of system parameters
(including quite low temperatures), are especially remarkable since the model
is implemented with our rather crude quasi-classical approach:  Quantized
initial conditions are selected for the action-angle variables, but then the
dynamics are completely classical (i.e., generated by Hamilton's equations,
without any more sophisticated \emph{semi}classical input). One is accustomed
to quasi-classical treatments being semi-quantitative, e.g., for quantized
vibrational and rotational dofs --- and even for electronically nonadiabatic
transitions between two or three electronic states --- but the level of
accuracy seen in the present work is much better than that reported in such
cases. It may be that the dense set of electronic states that constitute the
left and right electrodes in the present molecular model leads to a rapid
decay of the initial factorized state to the steady state, and thus makes
the overall system behave more classically.

The present results encourage one to expand the treatment to more realistic
molecular models of such phenomena. It is obvious how to include nuclear
degrees of freedom in the classical molecular dynamics simulations along
with the classical model for the electronics dofs. Also, MW's classical
electronic model is able to include electron correlation, i.e., two
electron interactions in the second-quantized Hamiltonian of the type
\begin{equation}
  \sum_{i,j,k,l} \bra{ij}\ket{kl} \op{a}^\dag_i\op{a}^\dag_j\op{a}_l\op{a}_k
  ,
\end{equation}
which would be necessary to describe effects such as Coulomb blockade.
Though the form of this more general classical electronic model has been
given, it has never been applied to any nontrivial examples. It will be of
interest to see how well the approach will work for such more realistic
molecular models.

\section{Acknowledgments}
This work was supported by the US-Israel Binational Science Foundation, by
the FP7 Marie Curie IOF project HJSC, by the National Science Foundation
Grant No. CHE-0809073 and by the Director, Office of Science, Office of
Basic Energy Sciences, Chemical Sciences, Geosciences, and Biosciences
Division, U.S. Department of Energy under Contract No. DE-AC02-05CH11231. We
also acknowledge a generous allocation of supercomputing time from the
National Energy Research Scientific Computing Center (NERSC) and the use of
the Lawrencium computational cluster resource provided by the IT Division at
the Lawrence Berkeley National Laboratory. TL is grateful to the The
Center for Nanoscience and Nanotechnology at Tel Aviv University of a
doctoral fellowship.  GC is grateful to the Azrieli Foundation for the
award of an Azrieli Fellowship.  ER thanks the Miller Institute for
Basic Research in Science at UC Berkeley for partial financial support
via a Visiting Miller Professorship.


%


\begin{thebibliography}{51}%
\makeatletter
\providecommand \@ifxundefined [1]{%
 \@ifx{#1\undefined}
}%
\providecommand \@ifnum [1]{%
 \ifnum #1\expandafter \@firstoftwo
 \else \expandafter \@secondoftwo
 \fi
}%
\providecommand \@ifx [1]{%
 \ifx #1\expandafter \@firstoftwo
 \else \expandafter \@secondoftwo
 \fi
}%
\providecommand \natexlab [1]{#1}%
\providecommand \enquote  [1]{``#1''}%
\providecommand \bibnamefont  [1]{#1}%
\providecommand \bibfnamefont [1]{#1}%
\providecommand \citenamefont [1]{#1}%
\providecommand \href@noop [0]{\@secondoftwo}%
\providecommand \href [0]{\begingroup \@sanitize@url \@href}%
\providecommand \@href[1]{\@@startlink{#1}\@@href}%
\providecommand \@@href[1]{\endgroup#1\@@endlink}%
\providecommand \@sanitize@url [0]{\catcode `\\12\catcode `\$12\catcode
  `\&12\catcode `\#12\catcode `\^12\catcode `\_12\catcode `\%12\relax}%
\providecommand \@@startlink[1]{}%
\providecommand \@@endlink[0]{}%
\providecommand \url  [0]{\begingroup\@sanitize@url \@url }%
\providecommand \@url [1]{\endgroup\@href {#1}{\urlprefix }}%
\providecommand \urlprefix  [0]{URL }%
\providecommand \Eprint [0]{\href }%
\providecommand \doibase [0]{http://dx.doi.org/}%
\providecommand \selectlanguage [0]{\@gobble}%
\providecommand \bibinfo  [0]{\@secondoftwo}%
\providecommand \bibfield  [0]{\@secondoftwo}%
\providecommand \translation [1]{[#1]}%
\providecommand \BibitemOpen [0]{}%
\providecommand \bibitemStop [0]{}%
\providecommand \bibitemNoStop [0]{.\EOS\space}%
\providecommand \EOS [0]{\spacefactor3000\relax}%
\providecommand \BibitemShut  [1]{\csname bibitem#1\endcsname}%
\let\auto@bib@innerbib\@empty
\bibitem [{\citenamefont {Nitzan}\ and\ \citenamefont
  {Ratner}(2003)}]{Nitzan2003}%
  \BibitemOpen
  \bibfield  {author} {\bibinfo {author} {\bibfnamefont {A.}~\bibnamefont
  {Nitzan}}\ and\ \bibinfo {author} {\bibfnamefont {M.~A.}\ \bibnamefont
  {Ratner}},\ }\href@noop {} {\bibfield  {journal} {\bibinfo  {journal}
  {Science}\ }\textbf {\bibinfo {volume} {300}},\ \bibinfo {pages} {1384}
  (\bibinfo {year} {2003})}\BibitemShut {NoStop}%
\bibitem [{\citenamefont {Leggett}\ \emph {et~al.}(1987)\citenamefont
  {Leggett}, \citenamefont {Chakravarty}, \citenamefont {Dorsey}, \citenamefont
  {Fisher}, \citenamefont {Garg},\ and\ \citenamefont {Zwerger}}]{Leggett87}%
  \BibitemOpen
  \bibfield  {author} {\bibinfo {author} {\bibfnamefont {A.~J.}\ \bibnamefont
  {Leggett}}, \bibinfo {author} {\bibfnamefont {S.}~\bibnamefont
  {Chakravarty}}, \bibinfo {author} {\bibfnamefont {A.~T.}\ \bibnamefont
  {Dorsey}}, \bibinfo {author} {\bibfnamefont {M.}~\bibnamefont {Fisher}},
  \bibinfo {author} {\bibfnamefont {A.}~\bibnamefont {Garg}}, \ and\ \bibinfo
  {author} {\bibfnamefont {W.}~\bibnamefont {Zwerger}},\ }\href@noop {}
  {\bibfield  {journal} {\bibinfo  {journal} {Rev. Mod. Phys.}\ }\textbf
  {\bibinfo {volume} {59}},\ \bibinfo {pages} {1} (\bibinfo {year}
  {1987})}\BibitemShut {NoStop}%
\bibitem [{\citenamefont {Landauer}(1970)}]{Landauer70}%
  \BibitemOpen
  \bibfield  {author} {\bibinfo {author} {\bibfnamefont {R.}~\bibnamefont
  {Landauer}},\ }\href@noop {} {\bibfield  {journal} {\bibinfo  {journal}
  {Philos. Mag.}\ }\textbf {\bibinfo {volume} {21}},\ \bibinfo {pages} {863}
  (\bibinfo {year} {1970})}\BibitemShut {NoStop}%
\bibitem [{\citenamefont {B\"{u}ttiker}(1986)}]{Buttiker1986}%
  \BibitemOpen
  \bibfield  {author} {\bibinfo {author} {\bibfnamefont {M.}~\bibnamefont
  {B\"{u}ttiker}},\ }\href@noop {} {\bibfield  {journal} {\bibinfo  {journal}
  {Phys. Rev. Lett.}\ }\textbf {\bibinfo {volume} {57}},\ \bibinfo {pages}
  {1761} (\bibinfo {year} {1986})}\BibitemShut {NoStop}%
\bibitem [{\citenamefont {Langreth}\ and\ \citenamefont
  {Abrahams}(1981)}]{Langreth81}%
  \BibitemOpen
  \bibfield  {author} {\bibinfo {author} {\bibfnamefont {D.~C.}\ \bibnamefont
  {Langreth}}\ and\ \bibinfo {author} {\bibfnamefont {E.}~\bibnamefont
  {Abrahams}},\ }\href@noop {} {\bibfield  {journal} {\bibinfo  {journal}
  {Phys. Rev. B}\ }\textbf {\bibinfo {volume} {24}},\ \bibinfo {pages} {2978}
  (\bibinfo {year} {1981})}\BibitemShut {NoStop}%
\bibitem [{\citenamefont {Haug}\ and\ \citenamefont
  {Jauho}(1996)}]{Jauho_book}%
  \BibitemOpen
  \bibfield  {author} {\bibinfo {author} {\bibfnamefont {H.}~\bibnamefont
  {Haug}}\ and\ \bibinfo {author} {\bibfnamefont {A.~P.}\ \bibnamefont
  {Jauho}},\ }\href@noop {} {\emph {\bibinfo {title} {Quantum Kinetics in
  Transport and Optics of Semiconductors}}}\ (\bibinfo  {publisher}
  {Springer},\ \bibinfo {address} {Germany},\ \bibinfo {year}
  {1996})\BibitemShut {NoStop}%
\bibitem [{\citenamefont {Datta}(1995)}]{Datta_book}%
  \BibitemOpen
  \bibfield  {author} {\bibinfo {author} {\bibfnamefont {S.}~\bibnamefont
  {Datta}},\ }\href@noop {} {\emph {\bibinfo {title} {Electronic Transport in
  Mesoscopic Systems}}}\ (\bibinfo  {publisher} {Cambridge University Press},\
  \bibinfo {address} {Cambridge},\ \bibinfo {year} {1995})\BibitemShut
  {NoStop}%
\bibitem [{\citenamefont {White}(1992)}]{White92}%
  \BibitemOpen
  \bibfield  {author} {\bibinfo {author} {\bibfnamefont {S.~R.}\ \bibnamefont
  {White}},\ }\href@noop {} {\bibfield
  {journal} {\bibinfo  {journal} {Phys. Rev. Lett.}\ }\textbf {\bibinfo
  {volume} {69}},\ \bibinfo {pages} {2863} (\bibinfo {year}
  {1992})}\BibitemShut {NoStop}%
\bibitem [{\citenamefont {Schmitteckert}(2004)}]{Schmitteckert04}%
  \BibitemOpen
  \bibfield  {author} {\bibinfo {author} {\bibfnamefont {P.}~\bibnamefont
  {Schmitteckert}},\ }\href@noop {} {\bibfield
  {journal} {\bibinfo  {journal} {Phys. Rev. B}\ }\textbf {\bibinfo {volume}
  {70}},\ \bibinfo {pages} {121302} (\bibinfo {year} {2004})}\BibitemShut
  {NoStop}%
\bibitem [{\citenamefont {Anders}\ and\ \citenamefont
  {Schiller}(2005)}]{Anders05}%
  \BibitemOpen
  \bibfield  {author} {\bibinfo {author} {\bibfnamefont {F.~B.}\ \bibnamefont
  {Anders}}\ and\ \bibinfo {author} {\bibfnamefont {A.}~\bibnamefont
  {Schiller}},\ }\href@noop {} {\bibfield  {journal} {\bibinfo  {journal}
  {Phys. Rev. Lett.}\ }\textbf {\bibinfo {volume} {95}},\ \bibinfo {pages}
  {196801} (\bibinfo {year} {2005})}\BibitemShut {NoStop}%
\bibitem [{\citenamefont {M{\"u}hlbacher}\ and\ \citenamefont
  {Rabani}(2008)}]{Rabani2008}%
  \BibitemOpen
  \bibfield  {author} {\bibinfo {author} {\bibfnamefont {L.}~\bibnamefont
  {M{\"u}hlbacher}}\ and\ \bibinfo {author} {\bibfnamefont {E.}~\bibnamefont
  {Rabani}},\ }\href@noop {} {\bibfield  {journal} {\bibinfo  {journal} {Phys.
  Rev. Lett.}\ }\textbf {\bibinfo {volume} {100}},\ \bibinfo {pages} {176403}
  (\bibinfo {year} {2008})}\BibitemShut {NoStop}%
\bibitem [{\citenamefont {Weiss}\ \emph {et~al.}(2008)\citenamefont {Weiss},
  \citenamefont {Eckel}, \citenamefont {Thorwart},\ and\ \citenamefont
  {Egger}}]{Weiss08}%
  \BibitemOpen
  \bibfield  {author} {\bibinfo {author} {\bibfnamefont {S.}~\bibnamefont
  {Weiss}}, \bibinfo {author} {\bibfnamefont {J.}~\bibnamefont {Eckel}},
  \bibinfo {author} {\bibfnamefont {M.}~\bibnamefont {Thorwart}}, \ and\
  \bibinfo {author} {\bibfnamefont {R.}~\bibnamefont {Egger}},\ }\href@noop
  {}
   {\bibfield  {journal} {\bibinfo
  {journal} {Phys. Rev. B}\ }\textbf {\bibinfo {volume} {77}},\ \bibinfo
  {pages} {195316} (\bibinfo {year} {2008})}\BibitemShut {NoStop}%
\bibitem [{\citenamefont {Werner}, \citenamefont {Oka},\ and\ \citenamefont
  {Millis}(2009)}]{Werner09}%
  \BibitemOpen
  \bibfield  {author} {\bibinfo {author} {\bibfnamefont {P.}~\bibnamefont
  {Werner}}, \bibinfo {author} {\bibfnamefont {T.}~\bibnamefont {Oka}}, \ and\
  \bibinfo {author} {\bibfnamefont {A.~J.}\ \bibnamefont {Millis}},\
  }\href@noop {}
  {\bibfield  {journal} {\bibinfo
  {journal} {Phys. Rev. B}\ }\textbf {\bibinfo {volume} {79}},\ \bibinfo
  {pages} {035320} (\bibinfo {year} {2009})}\BibitemShut {NoStop}%
\bibitem [{\citenamefont {Schir\'o}\ and\ \citenamefont
  {Fabrizio}(2009)}]{Schiro09}%
  \BibitemOpen
  \bibfield  {author} {\bibinfo {author} {\bibfnamefont {M.}~\bibnamefont
  {Schir\'o}}\ and\ \bibinfo {author} {\bibfnamefont {M.}~\bibnamefont
  {Fabrizio}},\ }\href@noop {} {\bibfield
  {journal} {\bibinfo  {journal} {Phys. Rev. B}\ }\textbf {\bibinfo {volume}
  {79}},\ \bibinfo {pages} {153302} (\bibinfo {year} {2009})}\BibitemShut
  {NoStop}%
\bibitem [{\citenamefont {Segal}, \citenamefont {Millis},\ and\ \citenamefont
  {Reichman}(2010)}]{Segal10}%
  \BibitemOpen
  \bibfield  {author} {\bibinfo {author} {\bibfnamefont {D.}~\bibnamefont
  {Segal}}, \bibinfo {author} {\bibfnamefont {A.~J.}\ \bibnamefont {Millis}}, \
  and\ \bibinfo {author} {\bibfnamefont {D.~R.}\ \bibnamefont {Reichman}},\
  }\href@noop {} {\bibfield  {journal} {\bibinfo
  {journal} {Phys. Rev. B}\ }\textbf {\bibinfo {volume} {82}},\ \bibinfo
  {pages} {205323} (\bibinfo {year} {2010})}\BibitemShut {NoStop}%
\bibitem [{\citenamefont {Cohen}\ and\ \citenamefont {Rabani}()}]{Cohen11}%
  \BibitemOpen
  \bibfield  {author} {\bibinfo {author} {\bibfnamefont {G.}~\bibnamefont
  {Cohen}}\ and\ \bibinfo {author} {\bibfnamefont {E.}~\bibnamefont {Rabani}},\
  }\href@noop {} {\enquote {\bibinfo {title} {Nonequilibrium many-body
  projected dynamics at long time scales with real-time path integral monte
  carlo},}\ }\bibinfo {note} {In preparation}\BibitemShut {NoStop}%
\bibitem [{\citenamefont {Averin}\ and\ \citenamefont
  {Likharev}(1986)}]{Averin86}%
  \BibitemOpen
  \bibfield  {author} {\bibinfo {author} {\bibfnamefont {D.~V.}\ \bibnamefont
  {Averin}}\ and\ \bibinfo {author} {\bibfnamefont {K.~K.}\ \bibnamefont
  {Likharev}},\ }\href@noop {} {\bibfield  {journal} {\bibinfo  {journal} {J.
  Low Temp. Phys.}\ }\textbf {\bibinfo {volume} {62}},\ \bibinfo {pages} {345}
  (\bibinfo {year} {1986})}\BibitemShut {NoStop}%
\bibitem [{\citenamefont {Beenakker}(1991)}]{Beenakker91}%
  \BibitemOpen
  \bibfield  {author} {\bibinfo {author} {\bibfnamefont {C.~W.~J.}\
  \bibnamefont {Beenakker}},\ }\href@noop {} {\bibfield  {journal} {\bibinfo
  {journal} {Phys. Rev. B}\ }\textbf {\bibinfo {volume} {44}},\ \bibinfo
  {pages} {1646} (\bibinfo {year} {1991})}\BibitemShut {NoStop}%
\bibitem [{\citenamefont {Koch}\ and\ \citenamefont {{von
  Oppen}}(2005)}]{Koch05}%
  \BibitemOpen
  \bibfield  {author} {\bibinfo {author} {\bibfnamefont {J.}~\bibnamefont
  {Koch}}\ and\ \bibinfo {author} {\bibfnamefont {F.}~\bibnamefont {{von
  Oppen}}},\ }\href@noop {} {\bibfield  {journal} {\bibinfo  {journal} {Phys.
  Rev. Lett.}\ }\textbf {\bibinfo {volume} {94}},\ \bibinfo {pages} {206804}
  (\bibinfo {year} {2005})}\BibitemShut {NoStop}%
\bibitem [{\citenamefont {Meir}\ and\ \citenamefont {Wingreen}(1992)}]{Meir92}%
  \BibitemOpen
  \bibfield  {author} {\bibinfo {author} {\bibfnamefont {Y.}~\bibnamefont
  {Meir}}\ and\ \bibinfo {author} {\bibfnamefont {N.~S.}\ \bibnamefont
  {Wingreen}},\ }\href@noop {} {\bibfield  {journal} {\bibinfo  {journal}
  {Phys. Rev. Lett.}\ }\textbf {\bibinfo {volume} {68}},\ \bibinfo {pages}
  {2512} (\bibinfo {year} {1992})}\BibitemShut {NoStop}%
\bibitem [{\citenamefont {Werner}\ \emph {et~al.}(2010)\citenamefont {Werner},
  \citenamefont {Oka}, \citenamefont {Eckstein},\ and\ \citenamefont
  {Millis}}]{Werner10}%
  \BibitemOpen
  \bibfield  {author} {\bibinfo {author} {\bibfnamefont {P.}~\bibnamefont
  {Werner}}, \bibinfo {author} {\bibfnamefont {T.}~\bibnamefont {Oka}},
  \bibinfo {author} {\bibfnamefont {M.}~\bibnamefont {Eckstein}}, \ and\
  \bibinfo {author} {\bibfnamefont {A.~J.}\ \bibnamefont {Millis}},\
  }\href@noop {}
   {\bibfield  {journal} {\bibinfo
  {journal} {Phys. Rev. B}\ }\textbf {\bibinfo {volume} {81}},\ \bibinfo
  {pages} {035108} (\bibinfo {year} {2010})}\BibitemShut {NoStop}%
\bibitem [{\citenamefont {Galperin}, \citenamefont {Ratner},\ and\
  \citenamefont {Nitzan}(2007)}]{Nitzan07}%
  \BibitemOpen
  \bibfield  {author} {\bibinfo {author} {\bibfnamefont {M.}~\bibnamefont
  {Galperin}}, \bibinfo {author} {\bibfnamefont {M.~A.}\ \bibnamefont
  {Ratner}}, \ and\ \bibinfo {author} {\bibfnamefont {A.}~\bibnamefont
  {Nitzan}},\ }\href@noop {} {\bibfield  {journal} {\bibinfo  {journal} {J.
  Phys.: Condens. Matter}\ }\textbf {\bibinfo {volume} {19}},\ \bibinfo {pages}
  {103201} (\bibinfo {year} {2007})}\BibitemShut {NoStop}%
\bibitem [{\citenamefont {Tully}(1990)}]{Tully90}%
  \BibitemOpen
  \bibfield  {author} {\bibinfo {author} {\bibfnamefont {J.~C.}\ \bibnamefont
  {Tully}},\ }\href@noop {} {\bibfield  {journal} {\bibinfo  {journal} {J.
  Chem. Phys.}\ }\textbf {\bibinfo {volume} {93}},\ \bibinfo {pages} {1061}
  (\bibinfo {year} {1990})}\BibitemShut {NoStop}%
\bibitem [{\citenamefont {Webster}, \citenamefont {Rossky},\ and\ \citenamefont
  {Friesner}(1991)}]{Webster91a}%
  \BibitemOpen
  \bibfield  {author} {\bibinfo {author} {\bibfnamefont {F.~J.}\ \bibnamefont
  {Webster}}, \bibinfo {author} {\bibfnamefont {P.~J.}\ \bibnamefont {Rossky}},
  \ and\ \bibinfo {author} {\bibfnamefont {R.~A.}\ \bibnamefont {Friesner}},\
  }\href@noop {} {\bibfield  {journal} {\bibinfo  {journal} {Comput. Phys.
  Commun.}\ }\textbf {\bibinfo {volume} {63}},\ \bibinfo {pages} {494}
  (\bibinfo {year} {1991})}\BibitemShut {NoStop}%
\bibitem [{\citenamefont {Coker}\ and\ \citenamefont {Xiao}(1995)}]{Coker95}%
  \BibitemOpen
  \bibfield  {author} {\bibinfo {author} {\bibfnamefont {D.~F.}\ \bibnamefont
  {Coker}}\ and\ \bibinfo {author} {\bibfnamefont {L.}~\bibnamefont {Xiao}},\
  }\href@noop {} {\bibfield  {journal} {\bibinfo  {journal} {J. Chem. Phys.}\
  }\textbf {\bibinfo {volume} {102}},\ \bibinfo {pages} {496} (\bibinfo {year}
  {1995})}\BibitemShut {NoStop}%
\bibitem [{\citenamefont {Kapral}(2006)}]{Kapral06}%
  \BibitemOpen
  \bibfield  {author} {\bibinfo {author} {\bibfnamefont {R.}~\bibnamefont
  {Kapral}},\ }\href@noop {} {\bibfield  {journal} {\bibinfo  {journal} {Ann.
  Rev. Phys. Chem.}\ }\textbf {\bibinfo {volume} {57}},\ \bibinfo {pages} {129}
  (\bibinfo {year} {2006})}\BibitemShut {NoStop}%
\bibitem [{\citenamefont {Miller}(2001)}]{Miller2001}%
  \BibitemOpen
  \bibfield  {author} {\bibinfo {author} {\bibfnamefont {W.~H.}\ \bibnamefont
  {Miller}},\ }\href@noop {} {\bibfield  {journal} {\bibinfo  {journal} {J.
  Phys. Chem. A}\ }\textbf {\bibinfo {volume} {105}},\ \bibinfo {pages} {2942}
  (\bibinfo {year} {2001})}\BibitemShut {NoStop}%
\bibitem [{\citenamefont {Miller}(2006)}]{Miller06}%
  \BibitemOpen
  \bibfield  {author} {\bibinfo {author} {\bibfnamefont {W.~H.}\ \bibnamefont
  {Miller}},\ }\href@noop {} {\bibfield  {journal} {\bibinfo  {journal} {J.
  Chem. Phys.}\ }\textbf {\bibinfo {volume} {125}},\ \bibinfo {pages} {132305}
  (\bibinfo {year} {2006})}\BibitemShut {NoStop}%
\bibitem [{\citenamefont {Makri}(2004)}]{Makri04}%
  \BibitemOpen
  \bibfield  {author} {\bibinfo {author} {\bibfnamefont {N.}~\bibnamefont
  {Makri}},\ }\href@noop {} {\bibfield  {journal} {\bibinfo  {journal} {J.
  Phys. Chem. A}\ }\textbf {\bibinfo {volume} {108}},\ \bibinfo {pages} {806}
  (\bibinfo {year} {2004})}\BibitemShut {NoStop}%
\bibitem [{\citenamefont {Wang}, \citenamefont {Sun},\ and\ \citenamefont
  {Miller}(1998)}]{Miller98a}%
  \BibitemOpen
  \bibfield  {author} {\bibinfo {author} {\bibfnamefont {H.}~\bibnamefont
  {Wang}}, \bibinfo {author} {\bibfnamefont {X.}~\bibnamefont {Sun}}, \ and\
  \bibinfo {author} {\bibfnamefont {W.~H.}\ \bibnamefont {Miller}},\
  }\href@noop {} {\bibfield  {journal} {\bibinfo  {journal} {J. Chem. Phys.}\
  }\textbf {\bibinfo {volume} {108}},\ \bibinfo {pages} {9726} (\bibinfo {year}
  {1998})}\BibitemShut {NoStop}%
\bibitem [{\citenamefont {Makri}\ and\ \citenamefont
  {Thompson}(1998)}]{Makri98}%
  \BibitemOpen
  \bibfield  {author} {\bibinfo {author} {\bibfnamefont {N.}~\bibnamefont
  {Makri}}\ and\ \bibinfo {author} {\bibfnamefont {K.}~\bibnamefont
  {Thompson}},\ }\href@noop {} {\bibfield  {journal} {\bibinfo  {journal}
  {Chem. Phys. Lett.}\ }\textbf {\bibinfo {volume} {291}},\ \bibinfo {pages}
  {101} (\bibinfo {year} {1998})}\BibitemShut {NoStop}%
\bibitem [{\citenamefont {Wang}\ \emph {et~al.}(1999)\citenamefont {Wang},
  \citenamefont {Song}, \citenamefont {Chandler},\ and\ \citenamefont
  {Miller}}]{Miller99}%
  \BibitemOpen
  \bibfield  {author} {\bibinfo {author} {\bibfnamefont {H.}~\bibnamefont
  {Wang}}, \bibinfo {author} {\bibfnamefont {X.~Y.}\ \bibnamefont {Song}},
  \bibinfo {author} {\bibfnamefont {D.}~\bibnamefont {Chandler}}, \ and\
  \bibinfo {author} {\bibfnamefont {W.~H.}\ \bibnamefont {Miller}},\
  }\href@noop {} {\bibfield  {journal} {\bibinfo  {journal} {J. Chem. Phys.}\
  }\textbf {\bibinfo {volume} {110}},\ \bibinfo {pages} {4828} (\bibinfo {year}
  {1999})}\BibitemShut {NoStop}%
\bibitem [{\citenamefont {Thompson}\ and\ \citenamefont
  {Makri}(1999)}]{Makri99}%
  \BibitemOpen
  \bibfield  {author} {\bibinfo {author} {\bibfnamefont {K.}~\bibnamefont
  {Thompson}}\ and\ \bibinfo {author} {\bibfnamefont {N.}~\bibnamefont
  {Makri}},\ }\href@noop {} {\bibfield  {journal} {\bibinfo  {journal} {J.
  Chem. Phys.}\ }\textbf {\bibinfo {volume} {110}},\ \bibinfo {pages} {1343}
  (\bibinfo {year} {1999})}\BibitemShut {NoStop}%
\bibitem [{\citenamefont {Rabani}, \citenamefont {Egorov},\ and\ \citenamefont
  {Berne}(1999)}]{Rabani99c}%
  \BibitemOpen
  \bibfield  {author} {\bibinfo {author} {\bibfnamefont {E.}~\bibnamefont
  {Rabani}}, \bibinfo {author} {\bibfnamefont {S.~A.}\ \bibnamefont {Egorov}},
  \ and\ \bibinfo {author} {\bibfnamefont {B.~J.}\ \bibnamefont {Berne}},\
  }\href@noop {} {\bibfield  {journal} {\bibinfo  {journal} {J. Phys. Chem. A}\
  }\textbf {\bibinfo {volume} {103}},\ \bibinfo {pages} {9539} (\bibinfo {year}
  {1999})}\BibitemShut {NoStop}%
\bibitem [{\citenamefont {Wang}, \citenamefont {Thoss},\ and\ \citenamefont
  {Miller}(2000)}]{Miller00}%
  \BibitemOpen
  \bibfield  {author} {\bibinfo {author} {\bibfnamefont {H.}~\bibnamefont
  {Wang}}, \bibinfo {author} {\bibfnamefont {M.}~\bibnamefont {Thoss}}, \ and\
  \bibinfo {author} {\bibfnamefont {W.~H.}\ \bibnamefont {Miller}},\
  }\href@noop {} {\bibfield  {journal} {\bibinfo  {journal} {J. Chem. Phys.}\
  }\textbf {\bibinfo {volume} {112}},\ \bibinfo {pages} {47} (\bibinfo {year}
  {2000})}\BibitemShut {NoStop}%
\bibitem [{\citenamefont {Thoss}, \citenamefont {Wang},\ and\ \citenamefont
  {Miller}(2001)}]{Miller01b}%
  \BibitemOpen
  \bibfield  {author} {\bibinfo {author} {\bibfnamefont {M.}~\bibnamefont
  {Thoss}}, \bibinfo {author} {\bibfnamefont {H.}~\bibnamefont {Wang}}, \ and\
  \bibinfo {author} {\bibfnamefont {W.~H.}\ \bibnamefont {Miller}},\
  }\href@noop {} {\bibfield  {journal} {\bibinfo  {journal} {J. Chem. Phys.}\
  }\textbf {\bibinfo {volume} {114}},\ \bibinfo {pages} {9220} (\bibinfo {year}
  {2001})}\BibitemShut {NoStop}%
\bibitem [{\citenamefont {Nakayama}\ and\ \citenamefont
  {Makri}(2003)}]{Makri03}%
  \BibitemOpen
  \bibfield  {author} {\bibinfo {author} {\bibfnamefont {A.}~\bibnamefont
  {Nakayama}}\ and\ \bibinfo {author} {\bibfnamefont {N.}~\bibnamefont
  {Makri}},\ }\href@noop {} {\bibfield  {journal} {\bibinfo  {journal} {J.
  Chem. Phys.}\ }\textbf {\bibinfo {volume} {119}},\ \bibinfo {pages} {8592}
  (\bibinfo {year} {2003})}\BibitemShut {NoStop}%
\bibitem [{\citenamefont {Nakayama}\ and\ \citenamefont
  {Makri}(2005)}]{Makri05}%
  \BibitemOpen
  \bibfield  {author} {\bibinfo {author} {\bibfnamefont {A.}~\bibnamefont
  {Nakayama}}\ and\ \bibinfo {author} {\bibfnamefont {N.}~\bibnamefont
  {Makri}},\ }\href@noop {} {\bibfield  {journal} {\bibinfo  {journal} {Proc.
  Natl. Acad. Sci. USA}\ }\textbf {\bibinfo {volume} {102}},\ \bibinfo {pages}
  {4230} (\bibinfo {year} {2005})}\BibitemShut {NoStop}%
\bibitem [{\citenamefont {Martin-Fierro}\ and\ \citenamefont
  {Pollak}(2007)}]{Pollak07}%
  \BibitemOpen
  \bibfield  {author} {\bibinfo {author} {\bibfnamefont {E.}~\bibnamefont
  {Martin-Fierro}}\ and\ \bibinfo {author} {\bibfnamefont {E.}~\bibnamefont
  {Pollak}},\ }\href@noop {} {\bibfield  {journal} {\bibinfo  {journal} {J.
  Chem. Phys.}\ }\textbf {\bibinfo {volume} {126}},\ \bibinfo {pages} {164108}
  (\bibinfo {year} {2007})}\BibitemShut {NoStop}%
\bibitem [{\citenamefont {Liu}\ and\ \citenamefont {Miller}(2008)}]{Miller08a}%
  \BibitemOpen
  \bibfield  {author} {\bibinfo {author} {\bibfnamefont {J.}~\bibnamefont
  {Liu}}\ and\ \bibinfo {author} {\bibfnamefont {W.~H.}\ \bibnamefont
  {Miller}},\ }\href@noop {} {\bibfield  {journal} {\bibinfo  {journal} {J.
  Chem. Phys.}\ }\textbf {\bibinfo {volume} {128}},\ \bibinfo {pages} {144511}
  (\bibinfo {year} {2008})}\BibitemShut {NoStop}%
\bibitem [{\citenamefont {Moix}\ and\ \citenamefont {Pollak}(2008)}]{Pollak08}%
  \BibitemOpen
  \bibfield  {author} {\bibinfo {author} {\bibfnamefont {J.~M.}\ \bibnamefont
  {Moix}}\ and\ \bibinfo {author} {\bibfnamefont {E.}~\bibnamefont {Pollak}},\
  }\href@noop {} {\bibfield  {journal} {\bibinfo  {journal} {J. Chem. Phys.}\
  }\textbf {\bibinfo {volume} {129}},\ \bibinfo {pages} {064515} (\bibinfo
  {year} {2008})}\BibitemShut {NoStop}%
\bibitem [{\citenamefont {Egorov}, \citenamefont {Rabani},\ and\ \citenamefont
  {Berne}(1999)}]{Rabani99r}%
  \BibitemOpen
  \bibfield  {author} {\bibinfo {author} {\bibfnamefont {S.~A.}\ \bibnamefont
  {Egorov}}, \bibinfo {author} {\bibfnamefont {E.}~\bibnamefont {Rabani}}, \
  and\ \bibinfo {author} {\bibfnamefont {B.~J.}\ \bibnamefont {Berne}},\
  }\href@noop {} {\bibfield  {journal} {\bibinfo  {journal} {J. Phys. Chem. B}\
  }\textbf {\bibinfo {volume} {103}},\ \bibinfo {pages} {10978} (\bibinfo
  {year} {1999})}\BibitemShut {NoStop}%
\bibitem [{\citenamefont {Thoss}\ and\ \citenamefont {Wang}(2004)}]{Thoss2004}%
  \BibitemOpen
  \bibfield  {author} {\bibinfo {author} {\bibfnamefont {M.}~\bibnamefont
  {Thoss}}\ and\ \bibinfo {author} {\bibfnamefont {H.~B.}\ \bibnamefont
  {Wang}},\ }\href@noop {} {\bibfield  {journal} {\bibinfo  {journal} {Ann.
  Rev. Phys. Chem.}\ }\textbf {\bibinfo {volume} {55}},\ \bibinfo {pages} {299}
  (\bibinfo {year} {2004})}\BibitemShut {NoStop}%
\bibitem [{\citenamefont {Miller}\ and\ \citenamefont
  {White}(1986)}]{Miller86a}%
  \BibitemOpen
  \bibfield  {author} {\bibinfo {author} {\bibfnamefont {W.~H.}\ \bibnamefont
  {Miller}}\ and\ \bibinfo {author} {\bibfnamefont {K.~A.}\ \bibnamefont
  {White}},\ }\href@noop {} {\bibfield  {journal} {\bibinfo  {journal} {J.
  Chem. Phys.}\ }\textbf {\bibinfo {volume} {84}},\ \bibinfo {pages} {5059}
  (\bibinfo {year} {1986})}\BibitemShut {NoStop}%
\bibitem [{\citenamefont {Miller}\ and\ \citenamefont
  {McCurdy}(1978)}]{MillerMcCurdyClassicalElec}%
  \BibitemOpen
  \bibfield  {author} {\bibinfo {author} {\bibfnamefont {W.~H.}\ \bibnamefont
  {Miller}}\ and\ \bibinfo {author} {\bibfnamefont {C.~W.}\ \bibnamefont
  {McCurdy}},\ }\href@noop {} {\bibfield  {journal}
  {\bibinfo  {journal} {J. Chem. Phys}\ }\textbf {\bibinfo {volume} {69}},\
  \bibinfo {pages} {5163} (\bibinfo {year} {1978})}\BibitemShut {NoStop}%
\bibitem [{\citenamefont {McCurdy}, \citenamefont {Meyer},\ and\ \citenamefont
  {Miller}(1979)}]{Miller79a}%
  \BibitemOpen
  \bibfield  {author} {\bibinfo {author} {\bibfnamefont {C.~W.}\ \bibnamefont
  {McCurdy}}, \bibinfo {author} {\bibfnamefont {H.~D.}\ \bibnamefont {Meyer}},
  \ and\ \bibinfo {author} {\bibfnamefont {W.~H.}\ \bibnamefont {Miller}},\
  }\href@noop {} {\bibfield  {journal} {\bibinfo  {journal} {J. Chem. Phys.}\
  }\textbf {\bibinfo {volume} {70}},\ \bibinfo {pages} {3177} (\bibinfo {year}
  {1979})}\BibitemShut {NoStop}%
\bibitem [{\citenamefont {Meyer}\ and\ \citenamefont {Miller}(1979)}]{SMM}%
  \BibitemOpen
  \bibfield  {author} {\bibinfo {author} {\bibfnamefont {H.}~\bibnamefont
  {Meyer}}\ and\ \bibinfo {author} {\bibfnamefont {W.~H.}\ \bibnamefont
  {Miller}},\ }\href@noop {\bibfield  {journal}
  {\bibinfo  {journal} {J. Chem. Phys}\ }\textbf {\bibinfo {volume} {71}},\
  \bibinfo {pages} {2156} (\bibinfo {year} {1979})}\BibitemShut {NoStop}%
\bibitem [{\citenamefont {Meyer}\ and\ \citenamefont
  {Miller}(1980)}]{SMMvsCPP}%
  \BibitemOpen
  \bibfield  {author} {\bibinfo {author} {\bibfnamefont {H.}~\bibnamefont
  {Meyer}}\ and\ \bibinfo {author} {\bibfnamefont {W.~H.}\ \bibnamefont
  {Miller}},\ }\href@noop {} {\bibfield  {journal}
  {\bibinfo  {journal} {J. Chem. Phys}\ }\textbf {\bibinfo {volume} {72}},\
  \bibinfo {pages} {2272} (\bibinfo {year} {1980})}\BibitemShut {NoStop}%
\bibitem [{\citenamefont {Hod}, \citenamefont {Baer},\ and\ \citenamefont
  {Rabani}(2008)}]{Rabani08}%
  \BibitemOpen
  \bibfield  {author} {\bibinfo {author} {\bibfnamefont {O.}~\bibnamefont
  {Hod}}, \bibinfo {author} {\bibfnamefont {R.}~\bibnamefont {Baer}}, \ and\
  \bibinfo {author} {\bibfnamefont {E.}~\bibnamefont {Rabani}},\ }\href@noop {}
  {\bibfield  {journal} {\bibinfo  {journal} {J. Phys.: Cond. Mat.}\ }\textbf
  {\bibinfo {volume} {20}},\ \bibinfo {pages} {383201} (\bibinfo {year}
  {2008})}\BibitemShut {NoStop}%
\bibitem [{\citenamefont {Caspary}, \citenamefont {Berman},\ and\ \citenamefont
  {Peskin}(2003)}]{Peskin}%
  \BibitemOpen
  \bibfield  {author} {\bibinfo {author} {\bibfnamefont {M.}~\bibnamefont
  {Caspary}}, \bibinfo {author} {\bibfnamefont {L.}~\bibnamefont {Berman}}, \
  and\ \bibinfo {author} {\bibfnamefont {U.}~\bibnamefont {Peskin}},\
  }\href@noop {}
  {\bibfield  {journal} {\bibinfo
  {journal} {Chem. Phys. Lett}\ }\textbf {\bibinfo {volume} {369}},\ \bibinfo
  {pages} {232} (\bibinfo {year} {2003})}\BibitemShut {NoStop}%
\bibitem [{\citenamefont {Stock}(1995)}]{StockLanger}%
  \BibitemOpen
  \bibfield  {author} {\bibinfo {author} {\bibfnamefont {G.}~\bibnamefont
  {Stock}},\ }\href@noop {} {\bibfield  {journal} {\bibinfo
   {journal} {J. Chem. Phys.}\ }\textbf {\bibinfo {volume} {103}},\ \bibinfo
  {pages} {2888} (\bibinfo {year} {1995})}\BibitemShut {NoStop}%
\end{thebibliography}

\end{document}